\documentclass[journal]{IEEEtran}
\ifCLASSINFOpdf
   \usepackage[pdftex]{graphicx}
\else
  \usepackage[dvips]{graphicx}
\fi
\usepackage[cmex10]{amsmath}
\usepackage{amssymb}

\usepackage{caption}

\usepackage{tikz}
\usepackage{pgfplots}
\pgfplotsset{compat=1.3}
\usetikzlibrary{arrows,snakes,shapes}

\DeclareMathAlphabet{\mathbit}{OML}{cmr}{bx}{it}

\DeclareMathOperator{\Q}{Q}
\DeclareMathOperator{\E}{E}

\newcommand\Sign{\operatorname{sign}}

\DeclareMathOperator{\T}{\operatorname{T}}

\DeclareMathOperator{\fieldR}{\mathbb{R}}

\newcommand{\ve}[1]{\boldsymbol{#1}}

\newcommand{\exdi}[2]{\E_{#1} \left[#2\right]}

\newcommand{\sign}[1]{\Sign{\left(#1\right)}}

\newcommand{\qfunc}[1]{\Q \left(#1\right)}

\tikzset{box_2/.style={rectangle,inner sep=0.5pt,fill=white, minimum height=0.1cm,minimum width=0.1cm,draw=white,thin}}

\begin{document}
%
\title{Asymptotic Parameter Tracking Performance\\with Measurement Data of 1-bit Resolution}
%
%
%
\author{Manuel~Stein, Alexander~K\"urzl, Amine~Mezghani and~Josef~A.~Nossek

\thanks{This work was supported by the Heinrich and Lotte M\"uhlfenzl Foundation and the German Federal Ministry for Economic Affairs and Energy (Grant 50NA1110).}
\thanks{The authors are with the Institute for Circuit Theory and Signal Processing, Technische Universit\"at M\"unchen, 80290 Germany
(e-mail: manuel.stein@tum.de, a.kuerzl@tum.de, amine.mezghani@tum.de, josef.a.nossek@tum.de).}
}

\maketitle

\begin{abstract}
The problem of signal parameter estimation and tracking with measurement data of low resolution is considered. In comparison to an ideal receiver with infinite receive resolution, the performance loss of a simplistic receiver with $1$-bit resolution is investigated. For the case where the measurement data is preprocessed by a symmetric hard-limiting device with $1$-bit output, it is well-understood that the performance for low SNR channel parameter estimation degrades moderately by $2/\pi$ ($-1.96$ dB). Here we show that the $1$-bit quantization loss can be significantly smaller if information about the temporal evolution of the channel parameters is taken into account in the form of a state-space model. By the analysis of a Bayesian bound for the achievable tracking performance, we attain the result that the quantization loss in dB is in general smaller by a factor of two if the channel evolution is slow. For the low SNR regime, this is equivalent to a reduced loss of $\sqrt{2/\pi}$ ($-0.98$ dB). By simulating non-linear filtering algorithms for a satellite-based ranging application and a UWB channel estimation problem, both with low-complexity $1$-bit analog-to-digital converter (ADC) at the receiver, we verify that the analytical characterization of the tracking error is accurate. This shows that the performance loss due to observations with low amplitude resolution can, in practice, be much less pronounced than indicated by classical results. Finally, we discuss the implication of the result for medium SNR applications like channel estimation in the context of mobile wireless communications.
\end{abstract}

\begin{IEEEkeywords}
parameter estimation, tracking, hard-limiter, 1-bit ADC, channel estimation, ranging
\end{IEEEkeywords}
\IEEEpeerreviewmaketitle
\section{Introduction}
\IEEEPARstart{W}{hen} analyzing parameter estimation methods and algorithms in the context of statistical signal processing, it is often assumed that the digital measurement data is available with high resolution. Therefore, quantization effects can be neglected in the underlying model and an ideal system with infinite amplitude resolution is usually assumed for the analytical characterization of the receiver. However, in practice, the hardware complexity and the power dissipation of the required ADC scales exponentially $\mathcal{O}(2^b)$ with the number of resolution bits $b$. Consequently, high resolution ADCs are expensive to build and are power consuming during system operation. Further, the speed of the temporal sampling process is limited when operating at high resolution \cite{Walden99}. A work-around to this unattractive property of high resolution signal processing systems is to adapt the estimation and tracking algorithms intentionally to measurements of low resolution. This allows us to use an ADC of low complexity, have a small production cost and moderate power consumption, or to perform sampling at high rates. In the extreme case, the conversion from the analog to the digital domain is performed by a symmetric hard-limiter, providing a digital measurement output with $1$-bit resolution. For such an ADC device, the circuit design becomes trivial. It can be realized by a single comparator element with zero threshold voltage. Further, this extreme approach has the advantage that low-level digital signal processing operations, which involve the binary receive data, can be carried out hardware-efficiently by using $1$-bit arithmetics. Nevertheless, due to the strong non-linearity, the conceptual simplicity of low-resolution analog-to-digital conversion comes with a significant performance loss. The focus of this work is to characterize the performance gap between a simplistic signal processing system with $1$-bit measurement data and an ideal receiver with infinite resolution in the context of signal parameter estimation and tracking.
\subsection{Related work}
An interesting and long-standing result in statistical signal processing with quantized receive data \cite{Bennett48} is, that for low SNR applications, the performance loss associated with $1$-bit hard-limiting is moderate with ${{2}/{\pi}}$ ($-1.96$ dB) \cite{Vleck66}. Due to the attractive simplicity of ADCs with $1$-bit amplitude resolution, a variety of works \cite{Curry70,Lok98,Madsen00,Mez10} have analyzed the loss associated with this non-linear operation in the context of signal parameter estimation. Focusing on the problem of reliable communication over a noisy channel, the work \cite{Dabeer06} establishes the theoretical limit of the transmission rate with a $1$-bit ADC at the receiver. Another line of works studies different methods aiming at the reduction of the $1$-bit quantization loss. In \cite{Gilbert93,Shamai94,Koch10,Zhang12}, the possibility to increase the temporal sampling rate with $1$-bit ADC is discussed in the context of communication theory, while \cite{Koch13} takes into account the optimization of the hard-limiting threshold. In \cite{Papa01}, the quantization threshold is adaptively adjusted, whereas \cite{Dabeer06_2} and \cite{Dabeer08} consider the method of dithering for signal parameter estimation from quantized data. In contrast, \cite{Zeitler12} analyzes the benefit of dithering strategies with feedback. Adding noise prior to the quantization operation and exploiting the effect of stochastic resonance is studied in \cite{Rousseau03}. \cite{Balkan10} proves that a constant quantization threshold maximizes the Fisher information measure and its Bayesian version. The work \cite{Mez12} reveals that noise correlation can be beneficial for the information flow (Shannon information measure) through highly non-linear ADC devices, while by means of an estimation theoretic approach (Fisher information measure), the discussions \cite{SteinICASSP13,SteinWSA13} and \cite{SteinWCL15} show how to exploit this effect for statistical signal processing tasks by an adjusted design of the analog radio front-end. In the context of non-linear filtering, \cite{Karlsson05} and \cite{Duan2008_2} study the effect of coarse receive signal quantization, while \cite{Clements72,Sviestins00} and \cite{Duan2008} propose algorithms for state estimation and tracking with quantized measurement data and analyze their performance. 
\subsection{Contribution}
Here we follow the idea of including additional side information about the evolution of the channel into the digital signal processing of measurement data with low amplitude resolution. Different technical applications like wireless communication, radar, sonar or satellite-based positioning require the continuous inference of channel parameters at the receiver. As this process is performed subsequently on measurement blocks of short duration and the channel in general follows basic physical principles, a stochastic model which describes the short-time temporal evolution of the channel parameters can be derived in many situations. Such a model forms an additional source of information which can be exploited within the digital part of the receiver at high internal resolution. We show that for signal processing systems, where the measurement data is acquired from a sampling device with low amplitude resolution, the embedding of available side information into the formulation of the estimation problem plays an important role. By an asymptotic performance analysis based on Bayesian bounds for signal parameter tracking \cite{Tichavsky98,Simandl01,Ristic04,Trees07}, we show that significant performance gains can be achieved for quantized receivers if a state-space model is incorporated into the estimation algorithm and tracking over subsequent blocks is performed. In contrast to preliminary works \cite{Karlsson05} and \cite{Duan2008_2}, on the subject of state estimation with quantized measurements, we carry out an asymptotic performance analysis under slow channel parameter evolution and obtain an explicit relative loss of $\sqrt{{2}/{\pi}}$ ($-0.98$ dB) in the low SNR regime. The analysis shows that in general, the performance gap $\chi$ between two signal processing systems established under a Fisher or Bayesian estimation perspective diminishes to $\sqrt{\chi}$ when analyzed in conjunction with a slow evolving state-space model. This corresponds to a reduction of the performance loss in dB by a factor of two, making the result particularly interesting for situations where the performance loss is pronounced (e.g. $1$-bit signal processing in the medium to high SNR regime). Analyzing the rate of convergence of the estimation error under slow evolution reveals that the duration of the transient phase of the tracking process increases accordingly by $\sqrt{\chi^{-1}}$. With Monte-Carlo simulations using particle filters for channel estimation tasks in the context of low SNR satellite-based ranging and UWB communication, we verify that the established results can be translated into signal processing applications.  In the beginning, we briefly review the Fisher and the Bayesian approach onto estimation without a state-space model and discuss the performance loss attained within these frameworks when operating with $1$-bit measurement data.
\section{Observation Model}
For the discussion, an amplified sensor signal
\begin{align}
y(t)=\gamma s(t;\theta(t))+{\eta}(t),
\end{align}
$y(t)\in\fieldR$, is assumed. The analog signal $y(t)$ consists of a deterministic transmit signal $s(t;\theta(t))\in\fieldR$, attenuated by factor $\gamma\in\fieldR$. The signal $s(t;\theta(t))$ is modulated by a parameter $\theta(t)\in\fieldR$, which evolves over time $t\in\fieldR$. White random noise ${\eta}(t)\in\fieldR$, due to an analog low-noise amplifier behind the receive sensor, distorts the receive signal in an additive way. The receive signal $y(t)$ is low-pass filtered to a one-sided bandwidth of $B$ and sampled with a rate of $f_s=2B=\frac{1}{T_s}$. In the $k$-th processing block of duration $NT_s$ we combine $N$ subsequent samples to an observation vector
\begin{align}
\ve{y}_k=\gamma \ve{s}(\theta_k)+\ve{\eta}_k
\end{align}
$\ve{y}_k, \ve{s}(\theta_k), \ve{\eta}_k\in\fieldR^N$, with the individual vector entries
\begin{align}
[\ve{y}_{k}]_n&=y((k{-}1)NT_s+(n-1)T_s)\notag\\
[\ve{s}(\theta_k)]_n&=s((k{-}1)NT_s+(n-1)T_s;\theta_k)\notag\\
\theta_k&=\theta((k{-}1)NT_s)\notag\\
[\ve{\eta}_{k}]_n&=\eta((k{-}1)NT_s+(n-1)T_s),
\end{align}
where $n=1,\ldots,N$. By following this model we assume that the temporal evolution of the channel parameter $\theta(t)$ is slow compared to the sampling process, so that we approximate the parameter $\theta_k$ to be constant within the $k$-th block. Note that this imposes no general restriction. In practice, the sampling rate $f_s$ or the block length $N$ can be chosen such that the assumption of a constant block parameter is fulfilled with sufficiently high accuracy. The temporal evolution of the parameter over subsequent blocks can then be described in the form of a transition probability function $p(\theta_k|\theta_{k\text{-}1})$ with an initial prior $p(\theta_0)$ modeling the uncertainty about the channel parameter at the beginning of the receive process. The noise samples $\ve{\eta}_k$ form a multivariate Gaussian random variable with the properties
\begin{align}
\exdi{\eta}{\ve{\eta}_k}&=\ve{0},\quad\quad \forall k, \notag\\
\exdi{\eta}{\ve{\eta}_k\ve{\eta}^T_k}&=\ve{I},\quad\quad \forall k,
\end{align}
such that the conditional probability of the receive signal $\ve{y}_k$ in the $k$-th block can be written
\begin{align}
p(\ve{y}_k|\theta_k)&=\frac{ 1 }{(2\pi)^{\frac{N}{2}} } {\rm e}^{-\frac{1}{2} \big(\ve{y}_k-\gamma\ve{s}(\theta_k)\big)^{\rm T} \big(\ve{y}_k-\gamma\ve{s}(\theta_k)\big) }\notag\\
&= \frac{ 1 }{(2\pi)^{\frac{N}{2}} } \prod_{n=1}^{N} {\rm e}^{-\frac{1}{2} \big([\ve{y}_{k}]_n-\gamma[\ve{s}(\theta_k)]_n\big)^2}.
\end{align}
In the following, in order to take into account a $1$-bit ADC at the receiver, the receive signal is considered to be exclusively available in the form
\begin{align}
\ve{r}_k=\sign{\ve{y}_k},\label{system:model:sign}
\end{align}
where $\sign{x}$ is the element-wise signum function with the definition
\begin{align}
\operatorname{sign}(x)=
\begin{cases}
+1& \text{if } x \geq 0\\
-1& \text{if } x < 0.
\end{cases}
\end{align}
After this hard-limiting operation, the conditional probability of each binary receive sample $[\ve{r}_k]_n$ is
\begin{align}
p([\ve{r}_k]_n=+1|\theta_k) &=\int_{-\gamma [\ve{s}(\theta_k)]_n}^{\infty}p_{\eta}([\ve{\eta}_{k}]_n) {\rm{d}} {[\ve{\eta}_{k}]_n}\notag\\
&=\qfunc{-\gamma [\ve{s}(\theta_k)]_n}\notag\\
&=1-\qfunc{\gamma [\ve{s}(\theta_k)]_n}
\end{align}
and
\begin{align}
p([\ve{r}_k]_n=-1|\theta_k) &=\int_{-\infty}^{-\gamma [\ve{s}(\theta_k)]_n}p_{\eta}([\ve{\eta}_{k}]_n) {\rm{d}}[\ve{\eta}_{k}]_n\notag\\
&=1-\qfunc{-\gamma [\ve{s}(\theta_k)]_n}\notag\\
&=\qfunc{\gamma [\ve{s}(\theta_k)]_n},
\end{align}
such that
\begin{align}
p(\ve{r}_k|\theta_k) &= \prod_{n=1}^{N} \big(1-\qfunc{\gamma [\ve{r}_k]_n [\ve{s}(\theta_k)]_n}\big)\notag\\
&= \prod_{n=1}^{N} \qfunc{-\gamma [\ve{r}_k]_n [\ve{s}(\theta_k)]_n},
\end{align}
with $\qfunc{x}$ being the Q-function
\begin{align}
\qfunc{x}=\frac{1}{\sqrt{2\pi}} \int_{x}^{\infty} \exp{\Big(-\frac{z^2}{2}\Big)} {\rm d} z.
\end{align}
The final task of the receiver is to calculate a block-wise estimate $\hat{\theta}(\ve{r}_k)$ from the receive signal $\ve{r}_k$. The quality of the estimate $\hat{\theta}(\ve{r}_k)$ is judged on the basis of a quadratic error
\begin{align}
\epsilon_k=\big(\hat{\theta}(\ve{r}_k)-\theta_k\big)^2.
\end{align}
\section{Hard-limiting Loss - Fisher Estimation}
First we discuss the problem under a Fisher theoretic perspective \cite{Kay93}. The parameter $\theta_k$ is considered to be deterministic but unknown. Further, each block is processed independently without taking into account the temporal evolution of the channel parameter $\theta_k$. In this case, the optimum block-wise inference procedure is the maximum likelihood estimator (MLE)
\begin{align}
\hat{\theta}_{\operatorname{ML}}(\ve{r}_k)=\arg\max_{\theta_k\in\Theta} p(\ve{r}_k|\theta_k).
\end{align}
As the estimator is unbiased and asymptotically efficient, the mean square error (MSE) of the estimator
\begin{align}
\operatorname{MSE}(\theta_k)=\exdi{\ve{r}_k|\theta_k}{\big(\hat{\theta}_{\operatorname{ML}}(\ve{r}_k)-\theta_k\big)^2}
\end{align}
reaches the theoretical limit, the so-called Cram\'er-Rao lower bound (CRLB)
\begin{align}
\operatorname{MSE}(\theta_k)\geq\frac{1}{F(\theta_k)}.
\end{align}
The Fisher information measure $F(\theta_k)$ is defined
\begin{align}
F(\theta_k)&=\exdi{\ve{r}_{k}|\theta_k}{ \bigg(\frac{\partial \ln{p(\ve{r}_k|\theta_k)}}{\partial\theta_k}\bigg)^2 }\notag\\
&= \sum_{n=1}^{N} \exdi{ [\ve{r}_k]_n |\theta_k}{ \bigg(\frac{\partial \ln{p([\ve{r}_k]_n|\theta_k)}}{\partial\theta_k}\bigg)^2 }\notag\\
&= \sum_{n=1}^{N}  \frac{\Big(\frac{\partial {p([\ve{r}_{k}]_n=+1|\theta_k)} }{\partial\theta_k}\Big)^2}{ p([\ve{r}_{k}]_n=+1|\theta_k) } +\sum_{n=1}^{N}  \frac{ \Big(\frac{\partial {p([\ve{r}_{k}]_n=-1|\theta_k)} }{\partial\theta_k}\Big)^2 }{ p([\ve{r}_{k}]_n=-1|\theta_k) } .
\label{measure:fisher:quantize}
\end{align}
With the derivatives of the conditional probability function
\begin{align}
\frac{\partial p([\ve{r}_{k}]_n=+1|\theta_k)}{ \partial \theta_k} &=\frac{\gamma}{\sqrt{2\pi}}  \bigg[\frac{\partial \ve{s}(\theta_k)}{\partial \theta_k} \bigg]_n {\rm e}^{- \frac{\gamma^2[\ve{s}(\theta_k)]_n^2}{2}}\notag\\
\frac{\partial p([\ve{r}_{k}]_n=-1|\theta_k)}{ \partial \theta_k} &=-\frac{\gamma}{\sqrt{2\pi}}  \bigg[\frac{\partial \ve{s}(\theta_k)}{\partial \theta_k} \bigg]_n {\rm e}^{- \frac{\gamma^2[\ve{s}(\theta_k)]_n^2}{2}},
\end{align}
the information measure is found to be given by
\begin{align}
F(\theta_k)&= \frac{\gamma^2}{2\pi} \sum_{n=1}^{N} \frac{  \Big[\frac{\partial \ve{s}(\theta_k)}{\partial \theta_k} \Big]_n^2 {\rm e}^{-\gamma^2 [\ve{s}(\theta_k)]_n^2}}{ \qfunc{\gamma [\ve{s}(\theta_k)]_n} \qfunc{-\gamma [\ve{s}(\theta_k)]_n}} .
\end{align}
As a performance reference for the non-linear $1$-bit receiver (\ref{system:model:sign}), we consider an ideal receiver which has access to the high resolution signal $\ve{y}_k$. For this kind of receive system, the Fisher information measure in the $k$-th block is found to be
\begin{align}
F_\infty(\theta_k)&=\exdi{\ve{y}_k|\theta_k}{ \bigg(\frac{\partial \ln{p(\ve{y}_k|\theta_k)}}{\partial\theta_k}\bigg)^2 }\notag\\
&=\gamma^2 \bigg(\frac{\partial \ve{s}(\theta_k)}{\partial \theta_k}\bigg)^{\T} \frac{\partial \ve{s}(\theta_k)}{\partial \theta_k}\notag\\
&=\gamma^2 \sum_{n=1}^{N}\bigg[\frac{\partial \ve{s}(\theta_k)}{\partial \theta_k} \bigg]_n^2.
\label{measure:fisher}
\end{align}
In order to compare both receivers, we define the relative performance loss by the block-wise information ratio
\begin{align}
\chi_k(\theta_k)=\frac{F(\theta_k)}{F_\infty(\theta_k)}.
\end{align}
As $\qfunc{0}=\frac{1}{2}$, we obtain
\begin{align}
\lim_{\kappa \to 0} \frac{ {\rm e}^{-\kappa^2} }{ \qfunc{\kappa} \qfunc{-\kappa}}=4
\end{align}
and the loss for asymptotically small SNR is 
\begin{align}
\lim_{\gamma \to 0}\chi_k(\theta_k)&=\frac{2}{\pi},\quad\quad \forall k.
\end{align}
\section{Hard-limiting Loss - Bayesian Estimation}
The Bayesian perspective is slightly different  \cite{Trees07}. Here the parameter $\theta_k$ is treated as a random variable which is distributed according to a block-wise prior $p(\theta_k)$. Still, each block is processed independently, but the prior knowledge $p(\theta_k)$ is incorporated into the estimation process. In such a situation, the optimum algorithm for the inference of the parameter $\theta_k$ is the conditional mean estimator (CME)
\begin{align}
\hat{\theta}_{\operatorname{CM}}(\ve{r}_k)&=\exdi{\theta_k|\ve{r}_k}{\theta_k}.
\end{align}
The MSE of any estimator for the Bayesian parameter estimation problem
\begin{align}
\operatorname{MSE}_k=\exdi{\ve{r}_k,\theta_k}{\big(\hat{\theta}(\ve{r}_k)-\theta_k\big)^2}
\end{align}
can be bounded by a Bayesian version of the CRLB
\begin{align}
\operatorname{MSE}_k\geq\frac{1}{J_k},
\end{align}
where the block-wise Bayesian information measure is
\begin{align}
J_k&=\exdi{\ve{r}_k,\theta_k}{ \bigg(\frac{\partial \ln p(\ve{r}_k,\theta_k)}{\partial \theta_k} \bigg)^2 }\notag\\
&=\exdi{\theta_k}{ \exdi{\ve{r}_k|\theta_k}{\bigg(\frac{\partial \ln p(\ve{r}_k|\theta_k)}{\partial \theta_k} \bigg)^2} }+\notag\\
&\hspace{0.41cm}\exdi{\theta_k}{ \bigg(\frac{\partial \ln p(\theta_k)}{\partial \theta_k} \bigg)^2 }\notag\\
&=\exdi{\theta_k}{ F(\theta_k) }+J_{p,k}\notag\\
&=\bar{F}_{k}+J_{p,k}.
\end{align}
Equivalently, for the ideal reference receiver we have 
\begin{align}
J_{\infty,k}&=\exdi{\ve{y}_k,\theta_k}{ \bigg(\frac{\partial \ln p(\ve{y}_k,\theta_k)}{\partial \theta_k} \bigg)^2 }\notag\\
&=\exdi{\theta_k}{ F_\infty(\theta_k) }+J_{p,k}\notag\\
&=\bar{F}_{\infty,k}+J_{p,k}.
\end{align}
Defining the relative performance gap between both systems
\begin{align}
\psi_k&=\frac{J_k}{J_{\infty,k}}\notag\\
&=\frac{ \bar{F}_{k} +J_{p,k} }{ \bar{F}_{\infty,k} +J_{p,k}},
\end{align}
allows us to compare both systems in the Bayesian setting. The case of technical interest is \begin{align}
J_{p,k} &\ll \bar{F}_{k},\quad\quad\forall k,\notag\\
J_{p,k} &\ll \bar{F}_{\infty,k},\quad\quad\forall k,
\end{align} 
for which the asymptotic analysis in the low SNR regime 
\begin{align}
\lim_{\gamma \to 0} \psi_k&\approx\lim_{\gamma \to 0}\frac{\bar{F}_{k} }{ \bar{F}_{\infty,k} }\notag\\
&=\frac{2}{\pi},\quad\quad\forall k,
\end{align}
produces approximately the same result as in the Fisher estimation framework.
\section{Hard-limiting Loss - Tracking}
Finally, we assume that the available stochastic model $p(\theta_{k}|\theta_{k\text{-}1})$, describing the temporal evolution of the channel parameter from one block to another, is taken into account in an optimum way. This allows us to perform parameter estimation with tracking over subsequent blocks and to calculate the current block estimate $\hat{\theta}_k$ based on the observations of the current block and all preceding blocks. We assume that the channel parameter $\theta_k$ evolves according to a stochastic model of first order (autoregressive model of order one)
\begin{align}
\theta_k = \alpha\theta_{k\text{-}1}+z_{k},\label{model:linear:evolution}
\end{align}
where $\alpha\in\fieldR$ and the innovation $z_{k}\in\fieldR$ is a Gaussian random variable with the properties
\begin{align}
\exdi{z_{k}}{z_{k}}&=0,\quad\quad\forall k,\notag\\
\exdi{z_{k}}{z_{k}^2}&=\sigma^2,\quad\quad \forall k,
\end{align}
such that the transition probability function is given by
\begin{align}
p(\theta_{k}|\theta_{k\text{-}1})=\frac{1}{\sqrt{2\pi}\sigma} {\rm e}^{-\frac{ (\theta_{k}-\alpha\theta_{k\text{-}1})^2}{2\sigma^2}}.
\end{align}
For the first block we assume an initial prior
\begin{align}
p(\theta_{0})=\frac{1}{\sqrt{2\pi}\sigma_0} {\rm e}^{-\frac{ (\theta_{0}-\mu_{0})^2 }{2\sigma_0^2} }.
\end{align}
Note that for such a state-space model, the mean and the variance of the parameter evolve according to
\begin{align}
\exdi{\theta_k}{\theta_k}&=\alpha^k \mu_{0}\\
\exdi{\theta_k}{ (\theta_k-\exdi{\theta_k}{\theta_k})^2 }&=\alpha^{2k} \sigma_0^2 +  \bigg( \sum_{i=1}^{k} \alpha^{2(k-i)} \bigg) \sigma^2.
\end{align}
In order to avoid divergence of the state-space variance, we restrict $\alpha$ to the range $0\leq\alpha<1$, such that
\begin{align}
\lim_{k\to\infty} \exdi{\theta_k}{\theta_k}&=0\\
\lim_{k\to\infty} \exdi{\theta_k}{ (\theta_k-\exdi{\theta_k}{\theta_k})^2 } &=\frac{1}{1-\alpha^2} \sigma^2.
\end{align}
The optimum estimator in such a setup is the CME with all past observation blocks
\begin{align}
\hat{\theta}_{\operatorname{CM}}(\ve{R}_k)&=\exdi{\theta_k|\ve{R}_k}{\theta_k}\notag\\
&=\int_{\Theta_k}\theta_k p(\theta_k|\ve{R}_k) {\rm d}\theta_k, \label{estimator:tracking:cme}
\end{align}
where the observation matrix
\begin{align}
\ve{R}_k=\begin{bmatrix} \ve{r}_k &\ve{r}_{k\text{-}1} &\ldots &\ve{r}_{1}\end{bmatrix}
\end{align}
contains the receive signals of all past blocks up to the $k$-th block.
The MSE of this estimator
\begin{align}
\operatorname{MSE}_k=\exdi{\ve{R}_k,\theta_k}{\big(\hat{\theta}_{\operatorname{CM}}(\ve{R}_k)-\theta_k\big)^2}
\end{align}
can be lower bounded by
\begin{align}
\operatorname{MSE}_k\geq\frac{1}{U_k},
\end{align}
where the tracking information measure \cite{Tichavsky98} in block $k$
\begin{align}
U_{k}=D^{22}_{k}-D^{21}_{k} (U_{k\text{-}1}+D^{11}_{k})^{-1} D^{12}_{k}
\end{align}
is calculated recursively with
\begin{align}
D^{11}_{k}
&=\exdi{\theta_{k\text{-}1},\theta_{k}}{ \bigg( \frac{\partial \ln p(\theta_{k}|\theta_{k\text{-}1})}{\partial \theta_{k\text{-}1}} \bigg)^2}\notag\\
&=\exdi{\theta_{k\text{-}1}}{ \exdi{\theta_{k}|\theta_{k\text{-}1}}{ \bigg( \frac{\partial \ln p(\theta_{k}|\theta_{k\text{-}1})}{\partial \theta_{k\text{-}1}} \bigg)^2 } }\\
D^{12}_{k}
&=\exdi{\theta_{k\text{-}1},\theta_{k}}{ \frac{\partial \ln p(\theta_{k}|\theta_{k\text{-}1})}{\partial \theta_{k\text{-}1} } \frac{\partial \ln p(\theta_{k}|\theta_{k\text{-}1})}{\partial \theta_{k}}}\notag\\
&=\exdi{\theta_{k\text{-}1}}{ \exdi{\theta_{k}|\theta_{k\text{-}1}}{ \frac{\partial \ln p(\theta_{k}|\theta_{k\text{-}1})}{\partial \theta_{k\text{-}1} } \frac{\partial \ln p(\theta_{k}|\theta_{k\text{-}1})}{\partial \theta_{k}}} }\notag\\
&=D^{21}_{k}\\
D^{22}_{k}
&= \exdi{\theta_{k\text{-}1},\theta_{k}}{ \bigg( \frac{\partial \ln p(\theta_{k}|\theta_{k\text{-}1})}{\partial \theta_{k}} \bigg)^2 }+\notag\\
&\hspace{0.41cm}\exdi{\theta_{k},\ve{r}_{k}}{ \bigg( \frac{\partial \ln p(\ve{r}_{k}|\theta_{k})}{\partial \theta_{k}} \bigg)^2 }\notag\\
&= \exdi{\theta_{k\text{-}1}}{ \exdi{\theta_{k}|\theta_{k\text{-}1}}{ \bigg( \frac{\partial \ln p(\theta_{k}|\theta_{k\text{-}1})}{\partial \theta_{k}} \bigg)^2 } }+\bar{F}_{k}.
\end{align}
With the state-space model (\ref{model:linear:evolution}), the required derivatives are
\begin{align}
\frac{\partial \ln p(\theta_{k}|\theta_{k\text{-}1})}{\partial \theta_{k\text{-}1}}&=\frac{(\theta_{k}-\alpha\theta_{k\text{-}1})\alpha}{\sigma^2}\notag\\
\frac{\partial \ln p(\theta_{k}|\theta_{k\text{-}1})}{\partial \theta_{k}}&=-\frac{(\theta_{k}-\alpha\theta_{k\text{-}1})}{\sigma^2},
\end{align}
such that
\begin{align}
\exdi{\theta_{k\text{-}1}}{ \exdi{\theta_{k}|\theta_{k\text{-}1}}{ \bigg( \frac{\partial \ln p(\theta_{k}|\theta_{k\text{-}1})}{\partial \theta_{k\text{-}1}} \bigg)^2 } }&=\frac{\alpha^2}{\sigma^2}\label{state-space:information}\\
\exdi{\theta_{k\text{-}1}}{ \exdi{\theta_{k}|\theta_{k\text{-}1}}{ \bigg( \frac{\partial \ln p(\theta_{k}|\theta_{k\text{-}1})}{\partial \theta_{k}} \bigg)^2 } }&=\frac{1}{\sigma^2}\\
\exdi{\theta_{k\text{-}1}}{ \exdi{\theta_{k}|\theta_{k\text{-}1}}{ \frac{\partial \ln p(\theta_{k}|\theta_{k\text{-}1})}{\partial \theta_{k\text{-}1} } \frac{\partial \ln p(\theta_{k}|\theta_{k\text{-}1})}{\partial \theta_{k}}} }&=-\frac{\alpha}{\sigma^2}.
\end{align}
Consequently, the recursive rule for the computation of the tracking information measure $U_{k}$ is given by
\begin{align}
U_{k}&=\frac{1}{\sigma^2} - \frac{\alpha^2}{\sigma^4} \bigg(U_{k\text{-}1}+\frac{\alpha^2}{\sigma^2}\bigg)^{-1}+\bar{F}_{k}\notag\\
&=\bigg(\sigma^2+\frac{\alpha^2}{ U_{k\text{-}1}}\bigg)^{-1}+\bar{F}_{k}, \label{recursion:1bit}
\end{align}
and accordingly for the ideal receiver (infinite resolution)
\begin{align}
U_{\infty,k}&=\bigg(\sigma^2+\frac{\alpha^2}{ U_{\infty,k\text{-}1}}\bigg)^{-1}+\bar{F}_{\infty,k},
\end{align}
where the initial value is
\begin{align}
U_{0}=U_{\infty,0}&=\frac{1}{\sigma_0^2}.
\end{align}
\subsection{Steady-state Tracking Performance}
After an initial transient phase, the tracking algorithm reaches a steady-state such that the estimation error saturates and
\begin{align}
U_{k}\approx U_{k\text{-}1}, \quad\quad \forall k>K_\lambda,\label{condition:steady}
\end{align}
where $K_\lambda$ defines the end of the transient phase. Therefore
\begin{align}
U&=\lim_{k\to\infty}U_{k}\notag\\
&=\frac{1-\alpha^2}{2\sigma^2} + \frac{ \bar{F} }{2} + \sqrt{ \bigg(\frac{1-\alpha^2}{2\sigma^2} + \frac{\bar{F}}{2} \bigg)^2 +  \frac{\alpha^2 \bar{F}}{\sigma^2}},\label{measure:steady:state}
\end{align}
where the expected steady-state Fisher information is
\begin{align}
\bar{F}&=\lim_{k\to\infty}\exdi{\theta_{k}}{ F(\theta_{k}) }.
\end{align}
The situation that the last term $\frac{\alpha^2 \bar{F}}{\sigma^2}$ in (\ref{measure:steady:state}) dominates the tracking information measure $U$ arises if the two conditions
\begin{align}
\bigg(\frac{1-\alpha^2}{2\sigma^2}\bigg)^2&\ll\frac{\alpha^2 \bar{F}}{\sigma^2}\label{condition:one}\\
\bigg(\frac{\bar{F}}{2} \bigg)^2&\ll\frac{\alpha^2 \bar{F}}{\sigma^2}\label{condition:two}
\end{align}
are fulfilled. The first condition (\ref{condition:one}) can be reformulated
\begin{align}
(1-\alpha^2)^2\ll \alpha^2  \sigma^2 \bar{F}\label{condition:one:simp}
\end{align}
and the second condition (\ref{condition:two}) can be stated as
\begin{align}
 \bar{F} \ll \frac{\alpha^2 }{\sigma^2}.\label{condition:two:fin}
\end{align}
Substituting (\ref{condition:two:fin}) into (\ref{condition:one:simp}), we get
\begin{align}
1-\alpha^2\ll  \alpha^2,\label{condition:one:fin}
\end{align}
which is satisfied if we set $\alpha$ close to one. Hence, if $\alpha$ is close to one (see eq. (\ref{condition:one:fin})) and the informative quality of the state-space model indicated by $\frac{\alpha^2}{\sigma^2}$ (see eq. (\ref{state-space:information})) is much higher than the expected steady-state Fisher information $ \bar{F}$ of the observation model (\ref{condition:two:fin}), the steady-state tracking information measure $U$ can be approximated by
\begin{align}
U&\approx \sqrt{ \frac{\alpha^2 \bar{F}}{\sigma^2}  }.\label{approx:slow}
\end{align}
For the comparison between the quantized receiver and the ideal system, we define the $1$-bit quantization loss for parameter estimation and tracking in the $k$-th block as
\begin{align}
\rho_k=\frac{U_{k}}{U_{\infty,k}},\label{information:ratio:tracking}
\end{align}
such that asymptotically
\begin{align}
\rho&=\lim_{k\to\infty}\rho_k\notag\\
&=\frac{U}{U_\infty},\label{information:ratio:tracking:asym}
\end{align}
where the steady-state tracking information measure $U_{\infty}$ for the ideal reference receiver is
\begin{align}
U_{\infty}=\frac{1-\alpha^2}{2\sigma^2} + \frac{ \bar{F}_{\infty} }{2} + \sqrt{ \bigg(\frac{1-\alpha^2}{2\sigma^2} + \frac{\bar{F}_{\infty}}{2} \bigg)^2 +  \frac{\alpha^2 \bar{F}_{\infty}}{\sigma^2}},
\end{align}
with the expected steady-state Fisher information
\begin{align}
\bar{F}_{\infty}&=\lim_{k\to\infty}\exdi{\theta_{k}}{ F_{\infty}(\theta_{k}) }.
\end{align}
Under the assumption that the state-space model has much higher informative value than the observation model independent of the form of the receiver, i.e.,
\begin{align}
\bar{F} &\ll \frac{\alpha^2}{\sigma^2}\label{condition:two:ex:sys1}\\
\bar{F}_{\infty} &\ll \frac{\alpha^2}{\sigma^2},\label{condition:two:ex:sys2}
\end{align}
it is possible to evaluate the loss for a slow temporal evolution of the channel parameter according to
\begin{align}
\lim_{\alpha \to 1} \rho&\approx\sqrt{ \frac{   \bar{F} }{ \bar{F}_{\infty} } }.\label{result:slow}
\end{align}
Note that as long as (\ref{condition:two:ex:sys1}) and (\ref{condition:two:ex:sys2}) are fulfilled, the result (\ref{result:slow}) holds in general, independent of the considered SNR regime. This implies that compared to the Fisher or the Bayesian approach, tracking the parameter with the use of a slow evolving state-space model leads to a $1$-bit quantization loss in dB which is smaller by a factor of two. With the result (\ref{result:slow}), we can make the explicit statement that for signal parameter estimation and tracking in the low SNR regime, the relative $1$-bit quantization loss is
\begin{align}
\lim_{\gamma \to 0}\lim_{\alpha \to 1} \rho\approx\sqrt{ \frac{2}{\pi} }.
\end{align}
\subsection{Convergence and Transient Phase Analysis}
In order to further analyze the behavior of the $1$-bit quantized system, we consider the convergence of the recursive information measure (\ref{recursion:1bit}). The goal is to determine the number of measurement blocks which are required to fulfill the steady-state condition (\ref{condition:steady}). To this end, we define a transient phase of quality $\lambda>1$ with duration
\begin{align}
{K}_{\lambda}=\inf \Big\{ k \geq 1 \Big| {|U_{k}-U|}\leq10^{-\lambda}{|U_{0}-U|} \Big\}.
\end{align}
The measure ${K}_{\lambda}$ characterizes the delay from the start of the tracking procedure to the steady-state entry point. The rate of convergence $\nu\in\fieldR$ of recursion (\ref{recursion:1bit}) is found by solving
\begin{align}
\lim_{k\to\infty} \frac{|U_{k}-U|}{|U_{k\text{-}1}-U|^\nu}=\xi
\end{align}
for $\nu$ with constant $\xi\in\fieldR,\xi<\infty$. As the derivative
\begin{align}
\frac{\partial U_{k}}{ \partial U_{k\text{-}1}}\bigg|_{U_{k\text{-}1}=U}&=\alpha^2 (\sigma^2 U + \alpha^2 )^{-2}\neq0,
\end{align}
we have $\nu=1$, i.e., the order of convergence is linear and
\begin{align}
\xi=\alpha^2 (\sigma^2 U + \alpha^2 )^{-2}.
\end{align}
With $|U_{k}-U| \approx \xi^k |U_{0}-U|$, the duration ${K}_{\lambda}$ is found to be approximately
\begin{align}
{K}_{\lambda}\approx-\frac{\lambda}{\log{\xi}}.
\end{align}
Assuming that the conditions (\ref{condition:two:ex:sys1}) and (\ref{condition:two:ex:sys2}) are satisfied and $\sqrt{\sigma^2 \bar{F}} + \alpha>1$, it is possible to use the approximation
\begin{align}
\xi&\approx\Big(\sqrt{\sigma^2 \bar{F}} + \alpha\Big)^{-2}.
\end{align}
In this case,
\begin{align}
{K}_{\lambda}&\approx\frac{\lambda}{2 \log \Big(\sqrt{\sigma^2 \bar{F}} + \alpha\Big)}\label{block:length}.
\end{align}
Specifying the additional relative delay $\Delta$ which is introduced with $1$-bit quantization by
\begin{align}
\Delta=\frac{{K}_{\lambda}}{{K}_{\infty,\lambda}},
\end{align}
where ${K}_{\infty,\lambda}$ is the duration of the transient phase for the ideal receive system, we find
\begin{align}
\Delta&\approx\frac{\log \Big(\sqrt{\sigma^2 \bar{F}_{\infty}} + \alpha\Big)}{ \log \Big(\sqrt{\sigma^2 \bar{F}} + \alpha\Big) }\notag\\
&\approx\sqrt{\frac{\bar{F}_{\infty}}{\bar{F}}}
\end{align}
for $\alpha \to 1$, independent of the choice of the steady-state accuracy $\lambda$. Further, with
\begin{align}
\left.\sqrt{\frac{\bar{F}_{\infty}}{\bar{F}}}\right|_{\gamma\to0}&=\sqrt{\frac{\pi}{2}}\notag\\
&\approx1.25
\end{align}
it can be concluded that with slow parameter evolution ($\alpha\to1$) and low SNR, the transient phase with the $1$-bit receiver takes approximately $25\%$ more time than with the ideal system.
\section{Application Examples}
\subsection{Satellite-based Positioning at Low SNR}
As an application, we consider a satellite-based ranging problem where a transmitter sends a known periodic signal of the form
\begin{align}
x(t)=\sum_{c=-\infty}^{\infty}[\ve{b}]_{(1+\operatorname{mod}(c,C))} g(t-c T_c).\label{signal:pulse:shaper}
\end{align}
The vector $\ve{b}$ is a binary sequence with $C$ symbols. Each symbol has a duration $T_c$ and $g(t)$ is the corresponding band-limited rectangular transmit pulse. A Doppler-compensated receiver observes an attenuated and delayed copy of the transmit signal
\begin{align}
y(t)&=\gamma s(t;\theta(t))+\eta(t)\notag\\
&=\gamma x(t-\theta(t))+\eta(t)
\label{signal:analog:gnss}
\end{align}
with additive white noise $\eta(t)$. By band-limiting and sampling the analog signal (\ref{signal:analog:gnss}), the ideal receiver attains the digital receive signal
\begin{align}
\ve{y}_k=\gamma \ve{s}(\theta_k)+\ve{\eta}_k,
\end{align}
while a low-cost $1$-bit version of the receiver operates exclusively on the basis of the signal sign
\begin{align}
\ve{r}_k&=\sign{\ve{y}_k}\notag\\
&=\sign{\gamma \ve{s}(\theta_k)+\ve{\eta}_k}.
\end{align}
The temporal evolution of the time-delay parameter $\theta_k$ can be approximated by
\begin{align}
\theta_k=\alpha\theta_{k\text{-}1}+z_k.
\end{align}
Note, that in this radio-based ranging example, $\alpha$ is related to the movement of transmitter and receiver. For simplicity, we assume that the state-space parameter $\alpha$ is constant over the considered amount of blocks and is known at the receiver. The receiver's task is to estimate the distance to the transmitter in each block $k$ by measuring the time-delay parameter $\hat{\theta}_k$. 
\subsubsection{Tracking with a non-linear filter}
Because the optimum estimator (\ref{estimator:tracking:cme}) is difficult to calculate in this situation we use a suboptimal non-linear filter \cite{Ristic04} for simulations. The particle filter is based on approximating the posterior
\begin{align}
p(\theta_k|\ve{R}_{k})&\approx\sum_{l=1}^{L} w_k^l\delta(\theta_k-\theta_k^l)\notag\\
&=\tilde{p}(\theta_k| \ve{R}_{k} ),
\end{align}
by $L$ particles $\theta_k^l$. The particle weights $w_k^l\geq0$ satisfy
\begin{align}
\sum_{l=1}^{L} w_k^l=1,
\end{align}
such that a block-wise estimate $\hat{\theta}_k$ can be calculated by
\begin{align}
\hat{\theta}_k=\sum_{l=1}^{L} w_k^l \theta_k^l.
\end{align}
Using the transitional probability function $p(\theta_k|\theta_{k\text{-}1})$ as the importance density, the particle weights are updated recursively
\begin{align}
\tilde{w}_k^l=w_{k\text{-}1}^l p(\ve{r}_k|\theta_k^l)
\end{align}
and normalized
\begin{align}
{w}_k^l=\frac{\tilde{w}_k^l}{\sum_{l=1}^{L} \tilde{w}_k^l}.
\end{align}
If the effective number of particles
\begin{align}
L_{{ \rm eff}} = \frac{1}{ \sum_{l=1}^{L} (w_k^l)^2 }
\end{align}
falls below a certain threshold $\kappa$, i.e.,
\begin{align}
L_{{ \rm eff}} \leq \kappa L,
\end{align}
a resampling step is performed by replacing the particles with sampling $L$ times from $\tilde{p}(\theta_k|\ve{R}_{k})$.
\subsubsection{Results}
For simulations, we use the signal of the $5$-th GPS satellite with $C=1023$, $T_c=\frac{1}{1.023 \text{ MHz}}$ and a rectangular transmit pulse $g(t)$ \cite{GPS_spec}.  According to the chip rate, the one-sided band-width of the receiver is set to $B=1.023$ MHz. The sampling rate is set to $f_s=2B$ and each block has the duration $NT_s=1$ ms, i.e., a block contains $N=2046$ samples. The signal-to-noise ratio is set to $\operatorname{SNR_{dB}}=-15.0$ dB. For the state-space model, we choose $\alpha=1-10^{-3}$ and $\sigma=10^{-3}$ and the initialization setup is $\mu_0=398.7342\cdot T_c$ and $\sigma_0=0.1\cdot T_c$. For $K=250$ blocks, we generate $100$ delay processes and run the non-linear filters with $L=100$ particles for each delay process $1000$ times with independent noise realizations, while the resampling threshold is set to $\kappa=0.66$.
%
%
\pgfplotsset{legend style={rounded corners=2pt,nodes=right}}
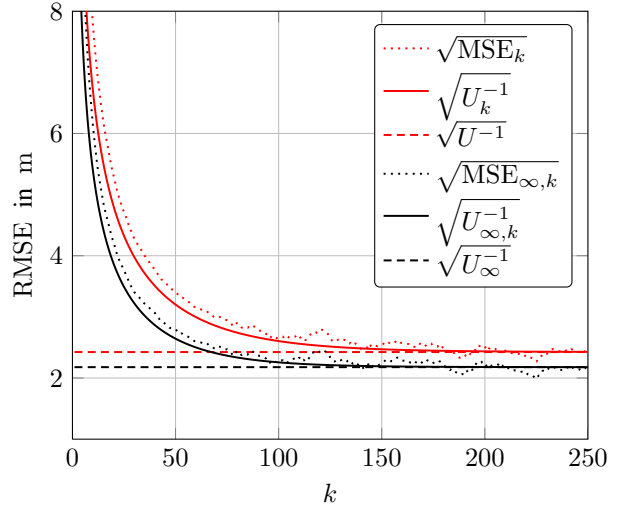
\begin{figure}[!htbp]
\begin{tikzpicture}[scale=1.0]

  	\begin{axis}[ylabel=$\operatorname{RMSE}\text{ in }\operatorname{m}$,
  			xlabel=$k$,
			grid,
			ymin=1,
			ymax=8,
			xmin=0,
			xmax=250,
			legend pos=north east]
			
			 \addplot[red, style=dotted, line width=0.75pt,smooth] table[x index=0, y index=2]{Tracking_Performance_Parallel_old.txt};
			 \addlegendentry{$\sqrt{ \operatorname{MSE}_k } $}
			 
			  \addplot[red, style=solid, line width=0.75pt,smooth] table[x index=0, y index=5]{Tracking_Performance.txt};
			  \addlegendentry{$\sqrt{U_k^{-1}}$}
			  
			  \addplot[red, style=densely dashed, line width=0.75pt,smooth] table[x index=0, y index=6]{Tracking_Performance.txt};
			 \addlegendentry{$\sqrt{U^{-1}}$}
			 
    			\addplot[black, style=dotted, line width=0.75pt,smooth] table[x index=0, y index=1]{Tracking_Performance_Parallel_old.txt};
			\addlegendentry{$\sqrt{\operatorname{MSE}_{\infty,k}}$}
			
			 \addplot[black, style=solid, line width=0.75pt,smooth] table[x index=0, y index=2]{Tracking_Performance.txt};
			 \addlegendentry{$\sqrt{U_{\infty,k}^{-1}}$}
			 
			 \addplot[black, style=densely dashed, line width=0.75pt,smooth] table[x index=0, y index=3]{Tracking_Performance.txt};
			 \addlegendentry{$\sqrt{U_{\infty}^{-1}}$}
			
	\end{axis}
	
\end{tikzpicture}
\caption{Tracking Error - Ranging}
\label{FigRE}
\end{figure}
The results depicted in Fig. \ref{FigRE} show that the block-wise analytic range tracking errors $U_k^{-1}$ and $U_{\infty,k}^{-1}$ in meter approach the asymptotic steady-state errors $U^{-1}$ and $U_{\infty}^{-1}$. Further, it can be observed that both non-linear filters are efficient, such that the errors $\operatorname{MSE}_k$ and $\operatorname{MSE}_{\infty,k}$ reach the theoretic tracking bounds $U_k^{-1}$ and $U_{\infty,k}^{-1}$. Therefore, in Fig. \ref{FigRL}, the quantization loss $\rho_k$ defined in (\ref{information:ratio:tracking}) is visualized. It is observed that at the beginning of the range tracking process, the performance gap between both receivers is moderate ($-1.38$ dB at $k=1$), due to the same initial knowledge with $\sigma_0^2$. In the transient phase, the quantization loss becomes quite pronounced ($-1.90$ dB at $k=15$). While reaching the steady-state phase ($k>250$), the loss converges to $-0.93$ dB.
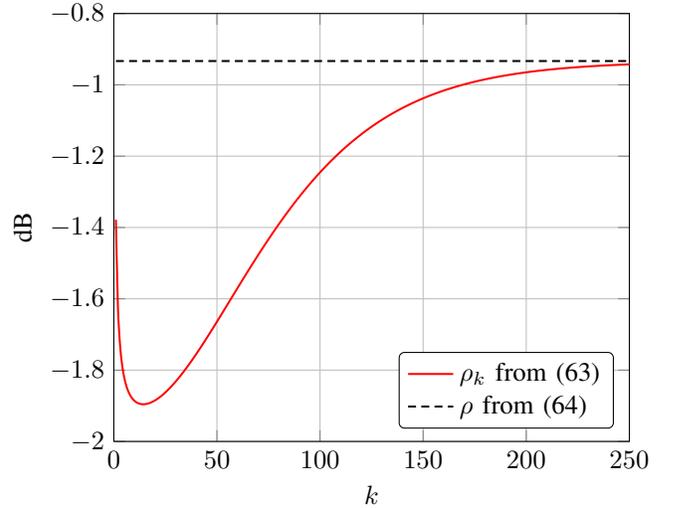
\begin{figure}[!htbp]
\begin{tikzpicture}[scale=1.0]

  	\begin{axis}[ylabel=$\text{dB}$,
  			xlabel=$k$,
			grid,
			ymin=-2.0,
			ymax=-0.8,
			xmin=0,
			xmax=250,
			legend pos=south east]
			
    			\addplot[red, style=solid, line width=0.75pt,smooth] table[x index=0, y index=8]{Tracking_Performance.txt};
			\addlegendentry{$\rho_k$ from (\ref{information:ratio:tracking})}
			
			\addplot[black, style=densely dashed, line width=0.75pt,smooth] table[x index=0, y index=9]{Tracking_Performance.txt};
			\addlegendentry{$\rho$ from (\ref{information:ratio:tracking:asym})}

	\end{axis}
	
\end{tikzpicture}
\caption{$1$-bit Tracking Loss - Ranging}
\label{FigRL}
\end{figure}
\subsection{UWB Channel Estimation at Low SNR}
For a second application we consider the estimation of the channel quality in the context of UWB communication. Similar to the ranging application, the receive signal of a synchronized receiver during a pilot phase can be modelled
\begin{align}
\ve{y}_k&= \ve{s}_k(\theta_k)+\ve{\eta}_k\notag\\
&=\theta_k \ve{x}_k+\ve{\eta}_k,\label{model:UWB:quality}
\end{align}
where $\ve{x}_k$ is the time-discrete form of a known unit power pilot signal with analog structure as in (\ref{signal:pulse:shaper}), and $\theta_k$ is the channel coefficient. Note, that in contrast to the ranging problem, the parameter $\theta_k$ in the ideal receive model (\ref{model:UWB:quality}) shows up in a linear form. The task of a low-cost $1$-bit UWB receiver 
\begin{align}
\ve{r}_k&=\sign{\ve{y}_k}\notag\\
&=\sign{\theta_k \ve{x}_{k}+\ve{\eta}_k}
\label{model:quant:UWB:quality}
\end{align}
is to estimate the signal attenuation $\hat{\theta}_k$ for each pilot block, while the channel coefficient follows the temporal evolution model (\ref{model:linear:evolution}). In contrast to the ranging application, we assume $B=528$ MHz, a Nyquist transmit pulse $g(t)$ of bandwidth $B$ and $C=10$ with $\operatorname{SNR_{dB}}=-15.0$ dB. The state-space model parameters are $\alpha=1-10^{-4}$ and $\sigma=\sqrt{(1-\alpha^2)\operatorname{SNR}}$,
where $\operatorname{SNR}=10^\frac{\operatorname{SNR_{dB}}}{10}$. The initialization setup is $\mu_0=\sqrt{\operatorname{SNR}}$ and $\sigma_0=0.05$.
\begin{figure}[!htbp]
\begin{tikzpicture}[scale=1.0]

  	\begin{axis}[ylabel=$\operatorname{RMSE}$,
  			xlabel=$k$,
			grid,
			ymin=0.015,
			ymax=0.045,
			xmin=0,
			xmax=250,
			legend pos=north east]
			
			\addplot[red, style=dotted, line width=0.75pt,smooth] table[x index=0, y index=4]{Tracking_Performance_CC_UWB.txt};
			 \addlegendentry{$\sqrt{ \operatorname{MSE}_k }$}
			 
			\addplot[red, style= solid, line width=0.75pt,smooth] table[x index=0, y index=5]{Tracking_Performance_CC_UWB.txt};
			\addlegendentry{$\sqrt{ U_k^{-1} }$}
			
			\addplot[red, style=dashed, line width=0.75pt,smooth] table[x index=0, y index=6]{Tracking_Performance_CC_UWB.txt};
			\addlegendentry{$\sqrt{ U^{-1} }$}
			
    			\addplot[black, style=dotted, line width=0.75pt,smooth] table[x index=0, y index=1]{Tracking_Performance_CC_UWB.txt};
			\addlegendentry{$\sqrt{ \operatorname{MSE}_{\infty,k} }$}
			
			\addplot[black, style=solid, line width=0.75pt,smooth] table[x index=0, y index=2]{Tracking_Performance_CC_UWB.txt};
			\addlegendentry{$\sqrt{ U_{\infty,k}^{-1} }$}
			
			\addplot[black, style=dashed, line width=0.75pt,smooth] table[x index=0, y index=3]{Tracking_Performance_CC_UWB.txt};
			\addlegendentry{$\sqrt{ U_{\infty}^{-1} }$}

	\end{axis}
	
\end{tikzpicture}
\caption{Tracking Error - UWB Channel Estimation}
\label{FigCC}
\end{figure}
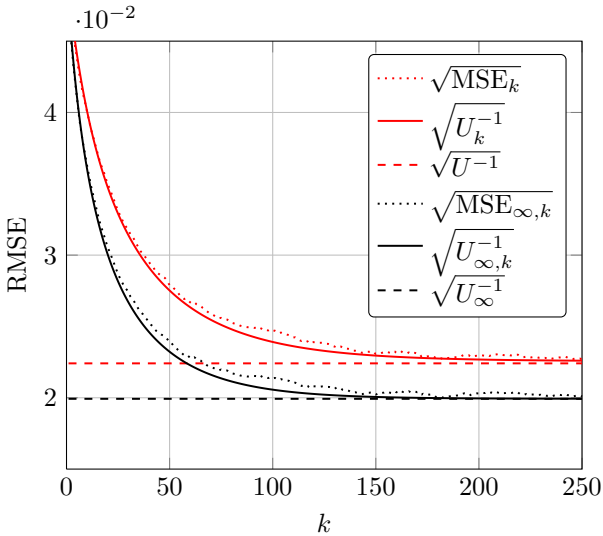
In Fig. \ref{FigCC} it can be seen, like in the ranging application, that the non-linear filters, simulated with $1000$ channel coefficient processes and $100$ independent noise realizations, perform efficiently and therefore closely to the tracking bounds $U_k^{-1}$ or $U_{\infty,k}^{-1}$. These bounds asymptotically equal the analytic steady-state errors $U^{-1}$ and $U_{\infty}^{-1}$.
\begin{figure}[!htbp]
\begin{tikzpicture}[scale=1.0]

  	\begin{axis}[ylabel=$\text{dB}$,
  			xlabel=$k$,
			grid,
			ymin=-1.5,
			ymax=-0.2,
			xmin=0,
			xmax=250,
			legend pos=north east]
			
    			\addplot[red, style=solid, line width=0.75pt,smooth] table[x index=0, y index=8]{Tracking_Performance_CC_UWB.txt};
			 \addlegendentry{$\rho_k$}
			 
			 \addplot[black, style=densely dashed, line width=0.75pt,smooth] table[x index=0, y index=9]{Tracking_Performance_CC_UWB.txt};
			 \addlegendentry{$\rho$}
			
	\end{axis}
	
\end{tikzpicture}
\caption{$1$-bit Tracking Loss - UWB Channel Estimation}
\label{FigCL}
\end{figure}
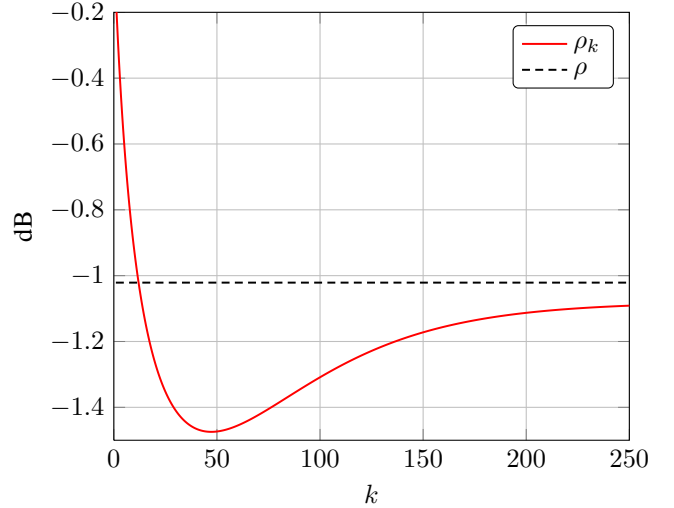
In Fig. \ref{FigCL}, the performance loss $\rho_k$ is depicted in dB. As in the ranging problem, it is observed that the loss after the initial transient phase recovers and approaches $-1.02$ dB. Note that for both of the considered applications, the asymptotic loss is slightly different from $-0.98$ dB, as the low SNR or the slow channel evolution assumptions are not fully valid for the chosen simulation setups.
\subsection{Enabling $1$-bit Estimation at Medium SNR}
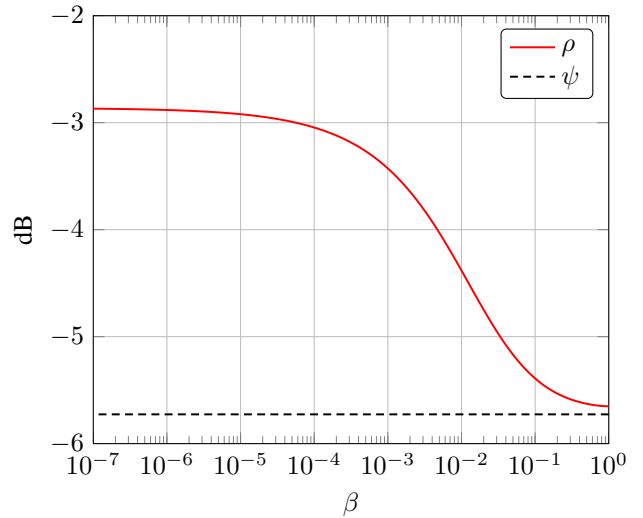
\begin{figure}[!htbp]
\begin{tikzpicture}[scale=1.0]

  	\begin{semilogxaxis}[ylabel=$\text{dB}$,
  			xlabel=$\beta$,
			grid,
			ymin=-6.0,
			ymax=-2.0,
			xmin=1e-7,
			xmax=1e-0,
			legend pos=north east]
			
    			\addplot[red, style=solid, line width=0.75pt,smooth] table[x index=0, y index=1]{Tracking_Performance_CC_MC_asym.txt};
			\addlegendentry{$\rho$}
			
			\addplot[black, style=densely dashed, line width=0.75pt,smooth] table[x index=0, y index=2]{Tracking_Performance_CC_MC_asym.txt};
			\addlegendentry{$\psi$}
			
\end{semilogxaxis}	
\end{tikzpicture}
\caption{$1$-bit Tracking Loss - Mobile Channel Estimation}
\label{FigCL_MC}
\end{figure}

Because a low-cost radio front-end design might be particularly interesting for mobile communication receivers, we finally investigate the potential tracking performance for signal parameter estimation in the medium SNR regime. As here the $1$-bit quantization loss is much more pronounced, using a low-cost $1$-bit ADC might make it impossible to meet the specified technical requirements. However, the use of a state-space model bears the potential to reduce the quantization loss, such that low-cost ADCs might become a possible system design option. For the considered scenario we assume a mobile communication channel as in (\ref{model:quant:UWB:quality}) with $B=2.5$ MHz, a pilot sequence of $C=10$ symbols and a medium channel quality of $\operatorname{SNR_{dB}}=6.0$ dB. The task of the receiver is to estimate the channel coefficient $\hat{\theta}_k$ in each pilot block, while the initial knowledge is assumed to be $\mu_0=\sqrt{\operatorname{SNR}}$ under the uncertainty
\begin{align}
\sigma_0=\Big(\sqrt{ \exdi{\theta_0}{F_\infty(\theta_0) } }\Big)^{-1}. 
\end{align}
The process noise is set to $\sigma=\sqrt{(1-\alpha^2)\operatorname{SNR}}$. In Fig. \ref{FigCL_MC}, the steady-state tracking loss $\rho$ is depicted for $\beta=1-\alpha$.
In comparison to the quantization loss $\psi$ without tracking, it can be seen that the quantization loss becomes smaller when $\alpha$ approaches one. However, note that the amount of blocks $K_\lambda$ that are required in order to reach the steady-state and achieve the small loss indicated by $\rho$ can become large. 
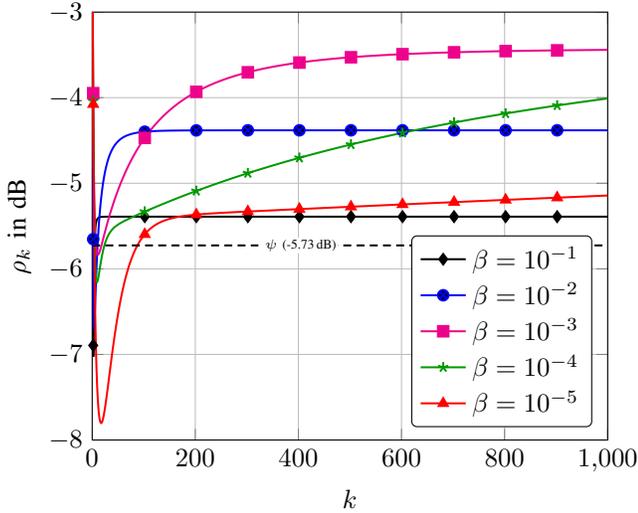
\begin{figure}[!htbp]
\begin{tikzpicture}[scale=1.0]

  	\begin{axis}[ylabel=${\rho_k}\text{ in dB}$,
  			xlabel=$k$,
			grid,
			ymin=-8,
			ymax=-3,
			xmin=0,
			xmax=1000,
			legend pos=south east]
			
    			\addplot[black, style=solid, line width=0.75pt,smooth, every mark/.append style={solid, fill=black}, mark=diamond*, mark repeat=100] table[x index=0, y index=5]{Tracking_Performance_CC_MC_1.txt};
			\addlegendentry{$\beta=10^{-1}$}
			
			\addplot[blue, style=solid, line width=0.75pt, smooth, every mark/.append style={solid, fill=black}, mark=otimes*, mark repeat=100] table[x index=0, y index=5]{Tracking_Performance_CC_MC_2.txt};
			\addlegendentry{$\beta=10^{-2}$}
			
    			\addplot[magenta, style=solid, line width=0.75pt,smooth,every mark/.append style={solid, fill=magenta}, mark=square*, mark repeat=100] table[x index=0, y index=5]{Tracking_Performance_CC_MC_3.txt};
			\addlegendentry{$\beta=10^{-3}$}
			
			\addplot[green!60!black, style=solid, line width=0.75pt,smooth,every mark/.append style={solid, fill=green!60!black}, mark=star, mark repeat=100] table[x index=0, y index=5]{Tracking_Performance_CC_MC_4.txt};
			\addlegendentry{$\beta=10^{-4}$}
			
    			\addplot[red, style=solid, line width=0.75pt,smooth,every mark/.append style={solid, fill=red}, mark=triangle*, mark repeat=100] table[x index=0, y index=5]{Tracking_Performance_CC_MC_5.txt};
			\addlegendentry{$\beta=10^{-5}$}

			\draw [densely dashed, line width=0.75pt]  (axis cs:0,-5.7262) --(axis cs:1000,-5.7262);
			\node[box_2] at(axis cs:400,-5.7262) {\tiny{ $\psi$ (-5.73\,dB) }};
			
	\end{axis}
	
\end{tikzpicture}
\caption{$1$-bit Tracking Loss - Mobile Channel Estimation}
\label{FigCL_MC_finite}
\end{figure}
In Fig. \ref{FigCL_MC_finite} the quantization loss $\rho_k$ for a finite amount of blocks and different $\beta$ is visualized. It becomes clear that in the considered scenario the reduction of the quantization error to a level above $-3.4$ dB ($\beta<10^{-3}$) might take a high number of blocks and therefore can only be realized with significant delay.
\section{Conclusion}
We have analyzed the performance gap between two extreme receive systems with respect to parameter estimation and tracking. The reference receiver performs analog-to-digital conversion with infinite amplitude resolution, while the low-cost receive system has a simple symmetric hard-limiting ADC with $1$-bit output resolution. If consecutive blocks are processed independently, we attain the well-established loss of $2/\pi$ ($-1.96$ dB) for low SNR applications. If, in contrast, additional side information about the temporal evolution of the channel in form of a state-space model is taken into account and the parameter is tracked over subsequent blocks, the loss can be significantly lower. For slow channel evolution ($\alpha\to1$), we attain $\sqrt{2/\pi}$ ($-0.98$ dB) in the low SNR regime, while for medium to high SNR, the loss in dB is, in general, smaller by a factor of two, compared to the case where the side information is not taken into account. Through simulation of a non-linear filtering algorithm we have verified that the result can be translated into practical applications. In particular, for situations with medium SNR, the result is interesting as here the quantization loss is pronounced. The embedding of additional information into the estimation and tracking process allows us to suppress the loss due to a non-linear radio front-end and therefore might enable new low-cost design options.
\section*{Acknowledgment}
The presented work was partially carried out during a research period of the second author at the National Chiao Tung University (NCTU) in Taiwan (August - October 2013). The authors would therefore like to express their gratitude to Prof. Po-Ning Chen and the Institute of Communications Engineering at NCTU Taiwan for hosting and support during this period. The support of the NCTU Taiwan Elite Internship Program is thankfully acknowledged. The authors would also like to thank the reviewers for their comments and suggestions, which helped to improve the quality of the article significantly. 
\end{document}